\documentclass[reprint,
 amsmath,amssymb,
 aps,
 prc,
]{revtex4-2}

\usepackage{graphicx}
\usepackage{dcolumn}
\usepackage{bm}
\usepackage{tikz}
\usepackage{quantikz}
\usepackage{setspace}
\usepackage{booktabs}
\usepackage{array}
\usepackage{tabularx}
\definecolor{linkpink}{RGB}{255,64,160}
\usepackage[colorlinks=true,linkcolor=blue,citecolor=blue,urlcolor=linkpink]{hyperref}

\begin{document}

\title{Quantum convolutional neural network for predicting nuclear charge radii}

\author{Jinzhe Wu}
\thanks{These authors contributed equally to this work.}
\affiliation{College of Physics, Jilin University, Changchun 130012, China}
\author{Jianping Zhao}
\thanks{These authors contributed equally to this work.}
\affiliation{College of Physics, Jilin University, Changchun 130012, China}
\author{Tianshuai Shang}
\affiliation{College of Physics, Jilin University, Changchun 130012, China}

\author{Yanhua Lu}
\affiliation{College of Physics, Jilin University, Changchun 130012, China}

\author{Yundong Wang}
\affiliation{College of Physics, Jilin University, Changchun 130012, China}

\author{Jian Li}
\email[E-mail: ]{jianli@jlu.edu.cn}
\affiliation{College of Physics, Jilin University, Changchun 130012, China}

\author{Haozhao Liang}
\affiliation{Department of Physics, Graduate School of Science,\\
The University of Tokyo, Tokyo 113-0033, Japan and\\
RIKEN Interdisciplinary Theoretical and Mathematical Sciences Program,
Wako 351-0198, Japan}
\date{\today}
\begin{abstract}
Quantum machine learning has the potential to become a new tool for understanding complex nuclear structures. In this work, we apply a hybrid quantum convolutional neural network (QCNN) to nuclear charge-radius prediction for the first time, aiming to explore the feasibility of quantum machine learning in nuclear-physics data analysis. Based on a classical convolutional neural network (CNN) framework, a small variational quantum convolutional filter is introduced as a quantum feature map to extract local correlations on the nuclear chart. The QCNN shows promising predictive accuracy and training stability, and provides a reliable description of the charge-radius evolution along several representative isotopic chains. These results support further investigation of quantum convolutional architectures for nuclear-structure data analysis.
\end{abstract}
\maketitle

\section{INTRODUCTION}
\label{sec:level1}

The nuclear charge radius characterizes the spatial distribution of nuclear charge and is a fundamental observable in nuclear-structure studies. The evolution of nuclear charge radii along isotopic chains can clearly reflect many nuclear structure phenomena, such as halo structures~\cite{nortershauser2008nuclear}, shape staggering and coexistence~\cite{yang2016isomer,marsh2018characterization},  nuclear magic numbers~\cite{kreim2014nuclear,gorges2019laser} and odd-even staggering (OES)~\cite{anselment1986odd,de2020measurement}. The nuclear charge radius can also be used to infer the thickness of the neutron skin, which is relevant to neutron-star properties such as radii and tidal deformabilities~\cite{fattoyev2018neutron,bano2023correlations}. In addition, high-precision nuclear charge-radius data provide important constraints for the development and testing of nuclear models~\cite{wang2013shell,reinhard2017toward,reinhard2021nuclear}. Therefore, more accurate predictions of nuclear charge radii are essential for gaining deeper insight into nuclear structure and related astrophysical problems.

Several experimental methods are currently available for measuring nuclear charge radii, including high-energy electron scattering from nuclei~\cite{de1987nuclear,kim1992ground}, atomic spectroscopy~\cite{heilig1974changes,aufmuth1987changes} and muonic-atom X-ray spectroscopy~\cite{engfer1974charge,fricke1995nuclear}. However, the available experimental data remain mainly limited to stable nuclei and long-lived unstable nuclei~\cite{angeli2013table,li2021compilation}. This motivates the development of more accurate and broadly applicable methods for predicting nuclear charge radii by combining theoretical models with experimental data.

Existing theoretical models for nuclear charge radii can be divided into two broad types: global and local models. Global models refer to theoretical frameworks or parameterized formulas that aim to provide a systematic description of the evolution of nuclear charge radii across the entire nuclear chart or broad nuclear regions. Representative examples include empirical formulas fitted to experimental data~\cite{nerlo1994simple,ru2009analysis}, Hartree-Fock models and Hartree-Fock-Bogoliubov models~\cite{stoitsov2003systematic,goriely2010further,nakada2019irregularities}, as well as relativistic mean-field models~\cite{lalazissis1999ground,zhao2010new,an2020novel}. Because these models aim at a systematic global description, their predictive accuracy for local charge-radius variations remains limited. In contrast, local models make use of correlations among neighboring nuclei on the nuclear chart, including Garvey-Kelson relations~\cite{piekarewicz2010garvey,sun2014new,sheng2015effective}. High accuracy can be achieved when experimental data for neighboring nuclei are abundant and the local trends are smooth. However, their performance may suffer when neighboring experimental data are sparse.

Addressing these limitations requires a flexible framework that can identify useful correlations in available nuclear data. Machine learning has been widely adopted in nuclear physics because it can extract nonlinear features from complex data and perform accurate regression and classification tasks~\cite{boehnlein2022colloquium}. Typical applications include the prediction of nuclear masses~\cite{utama2016nuclear,niu2018nuclear,lu2025nuclear}, excited states~\cite{lasseri2020taming,akkoyun2022estimations,wang2022study}, $\alpha$ decay~\cite{rodriguez2019alpha,banos2019bayesian}, $\beta$ decay~\cite{niu2019predictions,li2024comparative}, nuclear charge density~\cite{shang2022prediction,shang2024global,wang2026predictions}, magnetic moment~\cite{yuan2021magnetic}, nuclear level density~\cite{ozdougan2021estimations,du2024inference} and many-body quantum systems~\cite{carleo2017solving,wang2026neural}. For predicting nuclear charge radii, many machine-learning methods have been explored, including convolutional neural networks (CNN)~\cite{su2023progress,cao2023predictions}, support vector regression~\cite{jalili2024prediction}, naive Bayesian probability classifier~\cite{ma2020predictions}, Bayesian neural networks~\cite{autama2016nuclear,dong2022novel,dong2023nuclear}, artificial neural networks~\cite{akkoyun2013artificial,wu2020calculation,yang2023calibration}, kernel ridge regression~\cite{ma2022improved,tang2024nuclear}, decision-tree-based algorithms~\cite{Li2025Machine} and radial basis function approach~\cite{tao2023nuclear,li2026predictions,li2026extractions}. Among these approaches, CNNs are particularly suitable for charge-radius prediction because their convolutional filters can naturally exploit local charge-radius correlations among neighboring nuclei on the nuclear chart, as demonstrated by a representative CNN-based study that reduced the root-mean-square error (RMSE) to 0.0108 fm~\cite{su2023progress}.

Recent CNN-based studies raise the question of whether alternative local feature maps can improve charge-radius regression. With advances in quantum computing and hybrid quantum-classical algorithms, quantum machine learning has attracted increasing attention as a promising approach for constructing new feature representations and learning models~\cite{biamonte2017quantum,benedetti2019parameterized,havlicek2019supervised}. In particular, quantum convolutional neural network (QCNN) models introduce parameterized quantum circuits as local filters in a convolution-like structure. By encoding local classical inputs into quantum states and processing them through entangling operations and trainable quantum gates, such filters can generate nonlinear feature maps that differ from those produced by conventional filters. Previous studies have applied QCNN models to image classification~\cite{li2020quantum,henderson2020quanvolutional,liu2021hybrid,hur2022quantum,chen2023quantum,ceschini2025hybrid,sun2025scalable}. These works suggest that QCNN models may achieve improved performance over classical CNNs in certain settings~\cite{ceschini2025hybrid}. Moreover, because the same quantum convolutional filter can be reused across different local patches, the number of required qubits and the circuit depth can remain relatively small without requiring quantum random-access memory~\cite{henderson2020quanvolutional}. This structure is suitable for exploratory studies in the noisy intermediate-scale quantum (NISQ) era~\cite{preskill2018quantum}. In this context, quantum machine learning is commonly implemented through hybrid variational schemes~\cite{cerezo2021variational}, in which classical data are encoded into parameterized quantum circuits and the trainable circuit parameters are iteratively updated by a classical optimizer. 

This work follows this hybrid variational framework to examine whether a small quantum convolutional filter can enhance the learning of local nuclear-structure correlations~\cite{henderson2020quanvolutional}. We construct a QCNN for nuclear charge-radius prediction and compare it with a classical CNN. For each target nucleus, the local information on the nuclear chart is encoded as a five-channel tensor. To ensure a controlled comparison, both models share the same input representation, classical convolutional trunk, average-pooling layer (AvgPool), fully connected layers, and optimizer. The only architectural difference is that the QCNN replaces the final classical \(2\times2\) convolutional filter in the CNN with a weight-shared four-qubit variational quantum filter~\cite{liu2021hybrid}. This design allows the effect of introducing the quantum convolutional layer to be evaluated more directly while reducing possible influences from other model components or training settings. We evaluate both architectures in terms of overall accuracy and extrapolation performance on newly measured nuclei, and analyze their behavior through residual distributions and predictions along isotopic chains. The results show that the QCNN achieves a lower validation RMSE than the CNN and provides a better description of OES and shell structures along several representative isotopic chains. This behavior is consistent with the role of the variational quantum filter as a trainable nonlinear local feature map. By mapping each local patch into a quantum feature space before readout, the filter can encode multi-input correlations beyond the linear mapping of the replaced classical $2\times2$ convolutional filter~\cite{Schuld2019FeatureHilbertSpaces}. We expect this correlation-sensitive inductive bias to be useful for representing nuclear-chart patterns associated with OES and shell effects.

The paper is organized as follows. Section~\ref{sec:framework} presents the theoretical framework, including the data representation, model construction, and training method. Section~\ref{sec:results} presents and discusses the main results. Section~\ref{sec:discussion} summarizes the main conclusions and discusses future perspectives.

\begin{figure*}
\begin{tikzpicture}[
    >=Latex,
    font=\small,
    line join=round,
    line cap=round,
    flow/.style={-Latex, line width=1.0pt, draw=gray!72},
    conn/.style={line width=0.9pt, draw=gray!68},
    box/.style={draw=gray!60, rounded corners=2.5pt, line width=0.85pt, fill=gray!3, align=center},
    group/.style={draw=gray!45, rounded corners=3pt, dashed, dash pattern=on 3pt off 2.5pt, inner sep=6pt},
    titlebox/.style={font=\bfseries\small, align=center},
    blocktitle/.style={font=\bfseries\scriptsize, align=center},
    ann/.style={font=\scriptsize, inner sep=1pt, text=black, align=center},
    tinyann/.style={font=\tiny, inner sep=1pt, text=black, align=center},
    junction/.style={circle, fill=gray!70, inner sep=0.9pt}
]

\newcommand{\SmallGridFive}[4]{%
\begin{scope}[shift={(#1,#2)}]
  \pgfmathsetmacro{\cw}{#3/5}
  \foreach \i in {0,...,4}{
    \foreach \j in {0,...,4}{
      \draw[draw=gray!58, fill=#4!9] (\i*\cw,\j*\cw) rectangle ++(\cw,\cw);
    }
  }
\end{scope}}
\newcommand{\blocktitle}[1]{\textbf{\footnotesize #1}}
\newcommand{\FeatureStack}[7]{%
\begin{scope}[shift={(#2,#3)}]
  \foreach \i in {#6,...,1} {
    \pgfmathsetmacro{\dx}{#7*(\i-1)}
    \pgfmathsetmacro{\dy}{#7*(\i-1)}
    \draw[fill=gray!9, draw=gray!60] (\dx,\dy) rectangle ++(#4,#5);
    \SmallGridFive{0.0}{0}{#4}{gray}
  }
  \coordinate (#1-w) at (0,0.5*#5);
  \coordinate (#1-e) at ({0.10*(#6-1)+#4},{0.10*(#6-1)+0.5*#5});
\end{scope}}

\newcommand{\GridMap}[5]{%
\begin{scope}[shift={(#1,#2)}]
  \pgfmathsetmacro{\cw}{#3/#4}
  \pgfmathsetmacro{\ch}{#3/#5}
  \pgfmathtruncatemacro{\nxm}{#4-1}
  \pgfmathtruncatemacro{\nym}{#5-1}
  \draw[fill=gray!6, draw=gray!60] (0,0) rectangle (#3,{#5*\ch});
  \foreach \i in {1,...,\nxm}{\draw[gray!56] (\i*\cw,0)--(\i*\cw,{#5*\ch});}
  \foreach \j in {1,...,\nym}{\draw[gray!56] (0,\j*\ch)--(#3,\j*\ch);}
\end{scope}}

\begin{scope}[shift={(0.10,0.10)}]
  \def\W{2.65}
  \def\H{1.85}
  \def\sx{\W/12}
  \def\sy{\H/8}
  \foreach \i in {0,...,11}{
    \foreach \j in {0,...,7}{
      \draw[draw=gray!56, fill=gray!4] (\i*\sx,\j*\sy) rectangle ++(\sx,\sy);
    }
  }
  \foreach \i in {0,...,4}{
    \foreach \j in {3,...,7}{
      \draw[draw=gray!56, fill=blue!20] (\i*\sx,\j*\sy) rectangle ++(\sx,\sy);
    }
  }
  \draw[draw=gray!48, line width=0.9pt] (0,0) rectangle (\W,\H);
  \node[titlebox] at (1.32,2.2) {Nuclear chart};
  \node[ann, anchor=west] at (0.59,-0.3) {$5\times5$ Input};
\end{scope}
\draw[flow] (2.9,1) -- (4.1,1);
\FeatureStack{tensor}{4.3}{0.4}{0.70}{0.7}{5}{0.1}
\node[blocktitle] at (4.9,2.2) {Tensor};
\node[tinyann, scale=1.3] at (4.9,1.85) {$5\times5\times5$};
\node[ann] at (4.8,0) {$Z,N,V_Z,V_N,\Delta$};

\node[group, fit={(3.7,-0.2) (17,2.3)}, label={[titlebox]above:Shared classical trunk}] {};

\draw[flow] (5.8,1) -- (7,1);
\FeatureStack{conv1}{7.3}{0.3}{0.7}{0.7}{48}{0.01}
\node[blocktitle] at (6.3,1.6) {Conv 1};
\node[blocktitle] at (7.9,2.2) {Output1};
\node[tinyann, scale=1.3] at (6.3,1.3) {$3\times3$};
\node[tinyann, scale=1.3] at (7.9,1.85) {$48\times5\times5$};
\node[ann] at (6.3,0.6) {BN\\PReLU};

\draw[flow] (8.9,1) -- (10.1,1);
\FeatureStack{conv2}{10.4}{0.36}{0.7}{0.7}{36}{0.01}
\node[blocktitle] at (9.4,1.6) {Conv 2};
\node[tinyann, scale=1.3] at (9.4,1.3) {$3\times3$};
\node[blocktitle] at (10.9,2.2) {Output2};
\node[tinyann, scale=1.3] at (10.9,1.85) {$36\times5\times5$};
\node[ann] at (9.4,0.6) {BN\\PReLU};

\draw[flow] (11.8,1) -- (13,1);
\FeatureStack{conv3}{13.30}{0.64}{0.7}{0.7}{1}{0.1}
\node[blocktitle] at (12.3,1.6) {Conv 3};
\node[tinyann, scale=1.3] at (12.3,1.3) {$3\times3$};
\node[blocktitle] at (13.65,2.2) {Output3};
\node[tinyann, scale=1.3] at (13.65,1.85) {$1\times5\times5$};
\node[ann] at (12.3,0.6) {BN\\PReLU};
\node[group, fit={(0.5,-5.6) (6.7,-1.5)}, label={[titlebox]above:Alternative block},draw=none] {};
\draw[fill=green!4,dashed,draw=green!100] (0.3,-3.49) rectangle (6.9,-1.3);
\draw[fill=red!4,dashed,draw=red!100] (0.3,-3.51) rectangle (6.9,-5.8);
\draw[flow] (14.35,1) -- (15.55,1);
\begin{scope}[shift={(15.8,0.6)}]
  \draw[fill=gray!9, draw=gray!60] (0,0) rectangle (0.76,0.76);
  \foreach \i in {1,2} {\draw[gray!56] (\i*0.253,0)--(\i*0.253,0.76);}
  \foreach \j in {1,2} {\draw[gray!56] (0,\j*0.253)--(0.76,\j*0.253);}
\end{scope}
\node[blocktitle] at (14.92,1.6) {AvgPool};
\node[blocktitle] at (16.19,2.2) {Output4};
\node[tinyann, scale=1.3] at (16.19,1.85) {$1\times3\times3$};
\node[junction] (split) at (2,-3.5) {};

\draw[flow] (16.19,0.59) |- (16.19,-0.7) |- (0,-0.7) |-(0,-3.5) |- (split) |- (2,-2.35) --(2.6,-2.35);
\draw[flow] (split) |- (2,-4.65) --(2.6,-4.65);

\node[blocktitle, scale=1.7] at (1.3,-1.7) {CNN};
\node[box, scale=1.3,minimum width=2.45cm, minimum height=1cm,font=\footnotesize, fill=gray!6] (cnnbox) at (4.2,-2.35) {$2\times2$ \blocktitle{Conv 4}\\BN + PReLU};
\node[blocktitle, scale=1.7] at (1.3,-5.3) {QCNN};
\node[box, scale=1.5, minimum width=2.10cm, minimum height=1.2cm, fill=gray!6] (qcnnbox) at (4.2,-4.65) {};
\node[scale=0.5] at (4.18,-4.67) {
  \begin{quantikz}[row sep=0.20cm, column sep=0.25cm]
    \ket{0} & \gate{R_y} & \ctrl{1} & \qw & \qw & \targ{} & \gate{\mathrm{Rot}} & \meter{} \\
    \ket{0} & \gate{R_y} & \targ{} & \ctrl{1} & \qw & \qw & \gate{\mathrm{Rot}} & \qw \\
    \ket{0} & \gate{R_y} & \qw & \targ{} & \ctrl{1} & \qw & \gate{\mathrm{Rot}} & \qw \\
    \ket{0} & \gate{R_y} & \qw & \qw & \targ{} & \ctrl{-3} & \gate{\mathrm{Rot}} & \qw \\
  \end{quantikz}
};

\coordinate (split1) at (6.3,-3.5);
\fill[gray!70] (split1) circle (0.9pt);

\draw[flow] (5.8,-2.35) |- (6.3,-2.35) |- (split1) --(8.1,-3.5);
\draw[flow] (5.8,-4.65) |- (6.3,-4.65) |- (split1) --(8.1,-3.5);

\begin{scope}[shift={(8.2,-3.9)}]
  \GridMap{0}{0}{0.80}{2}{2}
\end{scope}
\node[blocktitle] at (8.6,-2.8) {Output5};
\node[ann] at (8.6,-4.3) {$1\times2\times2$};
\node[tinyann, scale=1.3] at (9.7,-3.2) {Flatten};
\node[tinyann, scale=1.3] at (9.7,-3.8) {PReLU};
\foreach \i in {0,1,...,3}
  \foreach \j in {0,1,...,9}
    \draw (10.5,-2-\i) --(12,-1.75-0.39*\j);
\foreach \j in {0,1,...,9}
  \draw (12,-1.7-0.4*\j) -- (13.5,-3.5);
\node[group, fit={(7.5,-5.6) (17,-1.5)}, label={[titlebox]above:Shared regression and evaluation}] {};

\draw[flow] (9.2,-3.5) -- (10.4,-3.5);
\foreach \k in {0,1,...,3}{
  \node[draw=gray!60, circle, minimum size=0.30cm, fill=gray!6] at (10.5,-2-\k) {};
}
\foreach \x in {0,1,...,9}
  \node[draw=gray!60, circle, minimum size=0.30cm, fill=gray!6]  at (12,-1.75-0.39*\x){};

\node[draw=gray!60, circle, minimum size=0.80cm, fill=gray!7] (out) at (13.5,-3.5) {$R_{\mathrm{th}}$};

\draw[flow] (14.05,-3.5) -- (14.9,-3.5);
\node[box, minimum width=2.10cm, minimum height=0.80cm, fill=gray!6] (evalbox) at (16,-3.5) {\textbf{Evaluation}\\RMSE};

\definecolor{optborder}{RGB}{235,190,95}
\definecolor{optfill}{RGB}{255,252,238}

\node[
  box,
  draw=optborder,
  fill=optfill,
  minimum width=2.95cm,
  minimum height=1.15cm
] (optbox) at (15.15,-5) {
  \textbf{Classical optimization}\\[1pt]
  $\mathcal{L}_{\mathrm{MSE}}(R_{\mathrm{th}},R_{\mathrm{exp}})$\\[1pt]
  RMSprop
};

\end{tikzpicture}
\caption{Architectures of the CNN and QCNN models adopted in this work. For each target nucleus, the input is a five-channel \(5\times5\) local window, including the proton number \(Z\), neutron number \(N\), shell-related features \(V_Z\) and \(V_N\), and the OES feature \(\Delta\). Both models share the first three convolutional layers and the average-pooling layer. The CNN then applies an additional \(2\times2\) convolutional layer, whereas the QCNN replaces this step with a four-qubit variational quantum filter. The resulting \(1\times2\times2\) output is flattened and passed through the shared fully connected layers to predict the charge radius. The MSE loss is minimized with RMSprop during training.}
\label{fig:cnn}
\end{figure*}
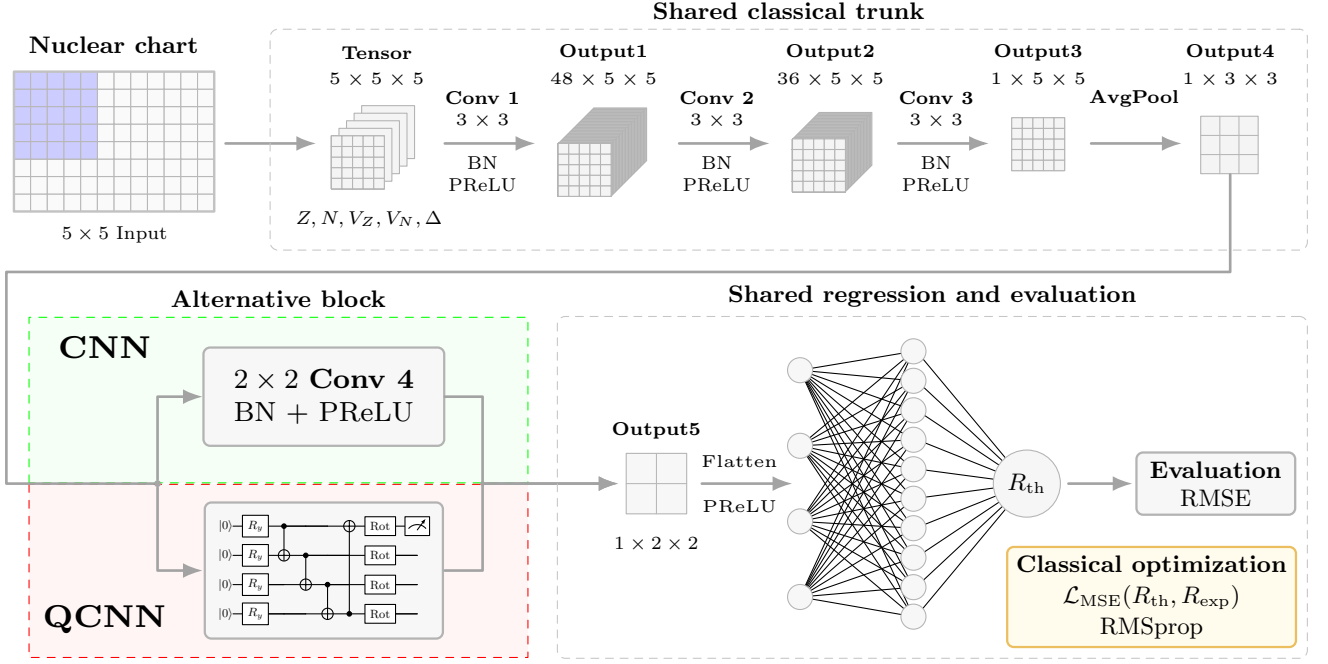

\section{Theoretical Framework}
\label{sec:framework}
This section describes the input features for nuclear charge-radius prediction, the CNN and QCNN architectures, and the training method. It also briefly describes the gradient calculation for the variational quantum circuit~\cite{mitarai2018quantum}, as well as the data sources and the construction of the extrapolation set.

Drawing inspiration from local charge-radius correlations on the nuclear chart~\cite{sun2014new}, we construct a local input tensor centered on each target nucleus. Specifically, the input is encoded as a \(5\times5\) local window on the nuclear chart. To construct the input channels, we use the proton number \(Z\) and neutron number \(N\) as two basic input features. We then introduce two additional features, $V_Z$ and $V_N$, defined as the distances from the nearest proton and neutron magic numbers not greater than the corresponding proton and neutron numbers, to encode shell-structure information. The proton and neutron magic-number sets are denoted by \(M_Z\) and \(M_N\), where \(M_Z=\{2,8,20,28,50,82\}\) and \(M_N=\{2,8,20,28,50,82,126\}\). The features \(V_Z\) and \(V_N\) are defined by
\begin{equation}
Z_M=\max\{m\in M_Z \mid m\le Z\}, V_Z = Z - Z_M.
\end{equation}
\begin{equation}
N_M=\max\{m\in M_N \mid m\le N\}, V_N = N - N_M.
\end{equation}

In addition, we introduce a fifth input feature, $\Delta$, to encode odd-even information associated with OES, which is closely related to pairing correlations~\cite{miller2019proton}. This feature is given by
\begin{equation}
\Delta=\frac{(-1)^Z+(-1)^N}{2}.
\end{equation}
Together, these five features form an input tensor of shape $5\times5\times5$.

The two architectures are shown in Fig.~\ref{fig:cnn}. In the CNN baseline, the first three convolutional layers use $3\times3$ filters with zero padding. Each layer is followed by batch normalization (BN)~\cite{ioffe2015batch} and PReLU~\cite{he2015delving} activation, with 48, 36, and 1 output channels. After these three convolutional layers, the input tensor is mapped to an intermediate feature tensor of shape $1\times5\times5$. This tensor is then reduced to $1\times3\times3$ by an adaptive AvgPool layer. The final convolutional layer uses a single $2\times2$ filter without zero padding. This layer is also followed by BN and PReLU activation, producing a final output tensor of shape $1\times2\times2$. This output is then flattened into a vector of length 4 and passed to a fully connected network with layer sizes of 4, 10, and 1. The output layer contains a single output node, which produces the predicted charge radius \(R_{\rm th}\).

The hybrid QCNN retains the same first three classical convolutional layers, AvgPool layer and fully connected layers as the CNN. The final classical $2\times2$ convolutional filter is replaced by a weight-shared four-qubit variational quantum filter. Four $2\times2$ local patches are extracted from the AvgPool output with stride 1 and no zero padding, and are fed sequentially into the variational quantum circuit. Here, weight sharing means that the same set of quantum circuit parameters is used for all local patches, analogous to using the same convolutional kernel at different spatial positions in a classical CNN~\cite{lecun1998gradient}. The structure of the variational quantum circuit is shown in Fig.~\ref{fig:q}. The circuit first encodes the four input values into four qubits initialized in the $\rvert 0\rangle$ state through $R_y$ gates~\cite{nielsen2010quantum}. It then applies four CNOT gates to entangle the qubits, followed by a trainable \(\mathrm{Rot}(\alpha,\beta,\gamma)\) gate~\cite{barenco1995elementary} on each qubit. Since each \(\mathrm{Rot}\) gate contains three rotation angles, the circuit contains 12 trainable rotation parameters in this implementation. Finally, the expectation value of the Pauli-$Z$ operator on the first qubit is used as the output of the quantum convolutional filter. Processing the four local patches yields four circuit outputs, which form a \(1\times2\times2\) tensor and are then passed through the same fully connected layers as in the CNN.

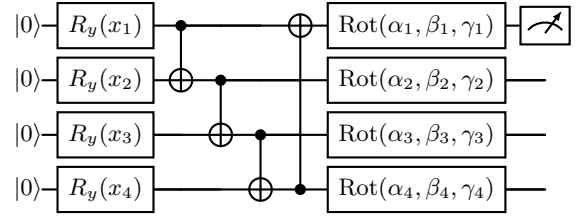
\begin{figure}
\small
\begin{quantikz}[row sep=0.125cm,column sep=0.20cm]
\ket{0} & 
\gate[][0.85cm][0.1cm]{R_y(x_1)} &  
\ctrl{1} & 
\qw & 
\qw & 
\targ{} & 
\gate{\mathrm{Rot}(\alpha_1, \beta_1, \gamma_1)} & 
\meter{} \\
\ket{0} & 
\gate{R_y(x_2)} & 
\targ{} & 
\ctrl{1} & 
\qw & 
\qw & 
\gate{\mathrm{Rot}(\alpha_2, \beta_2, \gamma_2)} & 
\qw \\
\ket{0} & 
\gate{R_y(x_3)} & 
\qw & 
\targ{} & 
\ctrl{1} & 
\qw & 
\gate{\mathrm{Rot}(\alpha_3, \beta_3, \gamma_3)} & 
\qw \\
\ket{0} & 
\gate{R_y(x_4)} & 
\qw & 
\qw & 
\targ{} & 
\ctrl{-3} & 
\gate{\mathrm{Rot}(\alpha_4, \beta_4, \gamma_4)} & 
\qw\\
\end{quantikz}
\caption{Variational quantum circuit construction for the quantum filter. The circuit comprises encoding, a variational block, and measurement. }
\label{fig:q}
\end{figure}

Both models are trained by minimizing the mean-squared-error (MSE) loss \(\mathcal{L}_{\rm MSE}\) between the predicted charge radius \(R_{\rm th}\) and the experimental value \(R_{\rm exp}\), using the RMSprop optimizer~\cite{ruder2016overview}. The optimizer hyperparameters are set as follows: the learning rate is \(\eta=1\times10^{-3}\), the RMSprop smoothing coefficient is \(\alpha_{\rm RMS}=0.99\), the numerical-stability constant is \(\epsilon=10^{-8}\), and the weight decay is set to 0.

The output of the variational quantum circuit can be written as an expectation value~\cite{chen2022quantum}:
\begin{equation}
f(x;\varTheta)
=
\langle \mathbf{0}|
U_0^{\dagger}(x)
U_q^{\dagger}(\varTheta)
\hat{A}
U_q(\varTheta)
U_0(x)
|\mathbf{0}\rangle .
\label{eq:parameter}
\end{equation}
Here, \(x=(x_1,x_2,x_3,x_4)\) denotes the flattened input of a local \(2\times2\) patch to the quantum filter, and \(|\mathbf{0}\rangle=|0\rangle^{\otimes 4}\) is the initial four-qubit state. The set of trainable circuit parameters is denoted by \(\varTheta=\{\theta_k\}\), where \(\theta_k\) is the \(k\)-th scalar parameter. The observable used in the final readout is denoted by \(\hat{A}\), which is the Pauli-\(Z\) operator on the first qubit in this work. In addition, \(U_0(x)\) denotes the input-encoding unitary operation implemented by the data-encoding gates, and \(U_q(\varTheta)\) denotes the post-encoding quantum-filter unitary, including the entangling gates and the trainable rotations.

Equation~\eqref{eq:parameter} also clarifies the feature-mapping role of the quantum filter. Defining \(|\psi_0(x)\rangle=U_0(x)|\mathbf{0}\rangle\) and \(\hat{O}(\varTheta)=U_q^\dagger(\varTheta)\hat{A}U_q(\varTheta)\), one obtains the equivalent form
\begin{equation}
f(x;\varTheta)
=
\langle \psi_0(x)|
\hat{O}(\varTheta)
|\psi_0(x)\rangle .
\end{equation}
Here, \(|\psi_0(x)\rangle\) is the four-qubit state after input encoding, and \(\hat{O}(\varTheta)\) is the effective observable determined by the quantum filter. Since four-qubit Pauli strings form an orthogonal basis for multi-qubit operators, \(\hat{O}(\varTheta)\) can be expanded as a linear combination of four-qubit Pauli strings~\cite{nielsen2010quantum}. For the \(R_y\) encoding, the \(i\)-th component of the local patch is encoded as
\begin{equation}
R_y(x_i)|0\rangle
=
\cos\frac{x_i}{2}|0\rangle
+
\sin\frac{x_i}{2}|1\rangle ,
\end{equation}
which gives \(\langle X_i\rangle=\sin x_i\), \(\langle Y_i\rangle=0\), and \(\langle Z_i\rangle=\cos x_i\) for a single encoded qubit. Since the encoded state is a product state, the expectation value of each Pauli string factorizes into a product of single-qubit expectation values. Combining these factors gives a finite real trigonometric representation of the quantum-filter output, consistent with
the Fourier and tensor-product feature representations of variational quantum models~\cite{schuld2021effect,shin2024dequantizing}:
\begin{equation}
f(x;\varTheta)
=
\sum_{\substack{S,T\subseteq\{1,2,3,4\}\\ S\cap T=\varnothing}}
C_{S,T}(\varTheta)
\prod_{i\in S}\sin x_i
\prod_{j\in T}\cos x_j .
\end{equation}
Here, \(S\) denotes the positions where the Pauli string contains \(X\), \(T\) denotes the positions where it contains \(Z\), and terms containing \(Y\) vanish for the \(R_y\) encoding. The effective coefficients \(C_{S,T}(\varTheta)\) arise from the Pauli-string expansion of \(\hat{O}(\varTheta)\) and depend on the circuit structure and parameters. The output is thus a trainable combination of trigonometric products, with each input \(x_i\) contributing a factor of \(1\), \(\sin x_i\), or \(\cos x_i\). By contrast, the replaced classical \(2\times2\) convolutional block first forms a linear local output \(z=w^{\mathrm T}x+b\), which is then normalized by BN and passed through PReLU activation. This provides a possible feature-mapping mechanism for the improved description of local nuclear-chart structures discussed in Sec.~\ref{sec:results}. 

The derivative of the quantum-filter output with respect to a trainable circuit parameter is computed with the parameter-shift rule, which can be written as~\cite{mitarai2018quantum}
\begin{equation}
\frac{\partial f(x;\varTheta)}{\partial \theta_k}
=
\frac{1}{2}
\left[
f\left(x;\varTheta_k^{+}\right)
-
f\left(x;\varTheta_k^{-}\right)
\right],
\label{eq:parameter_shift}
\end{equation}
where \(\varTheta_k^{\pm}\) denotes the same parameter set as \(\varTheta\), except that the \(k\)-th parameter is shifted as \(\theta_k\rightarrow\theta_k\pm\pi/2\).

The nuclear charge-radius data are taken from the CR2013 and CR2021 datasets~\cite{angeli2013table,li2021compilation}. Because neighboring-nucleus information is sparse for very light nuclei, nuclei satisfying both \(Z \leq 6\) and \(N \leq 6\) are excluded. The final dataset contains 1018 nuclei and is randomly split into training and validation sets with a ratio of 4:1. To examine the models' extrapolation performance, the 27 nuclei whose charge radii were measured after 2021 are selected as the extrapolation test set~\cite{geldhof2022impact,malbrunot2022nuclear,sommer2022charge,wang2024nuclear,konig2024nuclear,gustafsson2025charge}.

\section{results and discussion}
\label{sec:results}
In this section, we compare the performance of the CNN and QCNN introduced in Sec.~\ref{sec:framework}. For the single-run analyses below, we use the checkpoint with the lowest validation MSE for each model. The selected checkpoints occur at epochs 4718 for the CNN and 4413 for the QCNN. Table~\ref{tab:overall_rmse} summarizes the RMSE values for the training set, validation set, and full set. The QCNN yields lower RMSE values than the CNN on all three sets.

\begin{figure}
    \centering
    \setcounter{figure}{2}
    \includegraphics[width=1\linewidth]{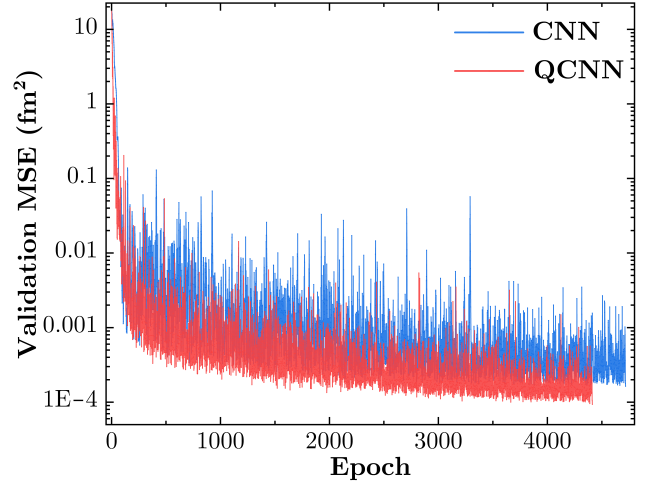}
    \caption{Validation MSE as a function of epoch for the CNN and QCNN, with the vertical axis plotted on a logarithmic scale. The blue and red curves denote the CNN and QCNN, respectively. The selected checkpoints correspond to the minima of the validation MSE curves, occurring at epochs 4718 and 4413 for the CNN and QCNN, respectively.}
    \label{fig:RMSE}
\end{figure}

Figure~\ref{fig:RMSE} shows the corresponding validation MSE curves. After a rapid decrease during the early stage of training, both models enter a slower optimization regime and subsequently fluctuate around a plateau. The QCNN reaches a lower loss level and exhibits smaller fluctuations than the CNN. Training beyond the selected checkpoints does not produce a sustained decrease in the validation MSE.

\begin{table}[!htb]
\renewcommand{\arraystretch}{1.1}
\centering
\caption{RMSE values of the CNN and QCNN on the training set, validation set, and full set.}
\label{tab:overall_rmse}
\vspace{1em} 
\setlength{\tabcolsep}{10pt}
\begin{tabular}{lccc}
\hline
\hline
Model & Train (fm) & Val (fm) & Full (fm) \\
\hline
CNN  & 0.0071 & 0.0124 & 0.0084 \\
QCNN & 0.0054 & 0.0100 & 0.0066 \\
\hline
\hline
\end{tabular}
\end{table}

\begin{figure}
    \centering
    \includegraphics[width=1\linewidth]{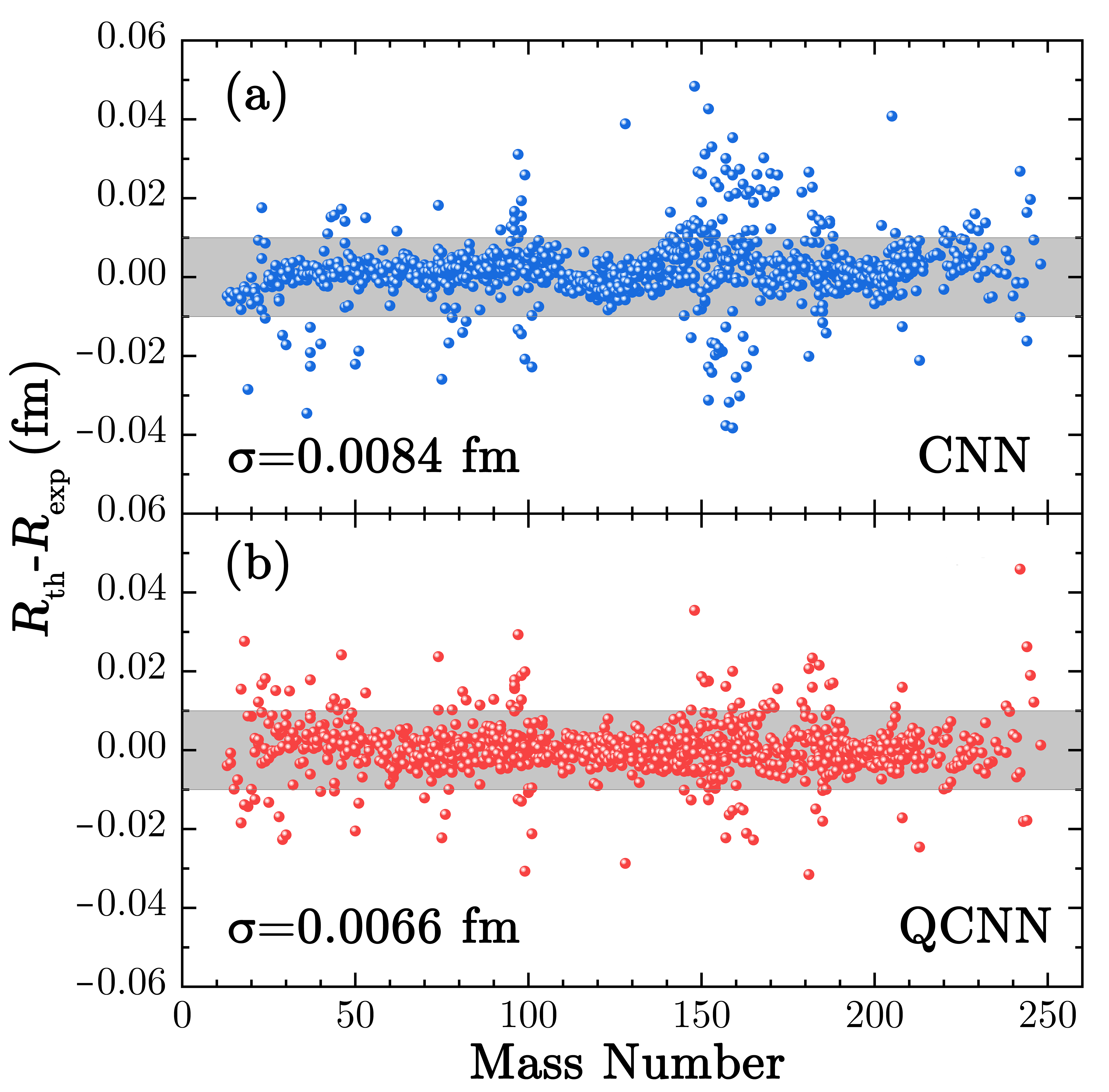}
    \caption{Residuals of the CNN and QCNN predictions relative to the experimental charge radii. Blue dots denote the CNN residuals, and red dots denote the QCNN residuals. The gray shaded area represents a range of \(\pm 0.01\) fm.}
    \label{fig:2}
\end{figure}

To examine how this overall error difference varies with mass number, Fig.~\ref{fig:2} presents the residuals of the CNN and QCNN predictions relative to the experimental charge radii. For most nuclei, the residuals of both models fall within \(\pm 0.01\) fm, although the CNN residuals are more dispersed. The residual distributions also vary across mass-number regions, as quantified in Table~\ref{tab:rmse_bins_N}. The largest reductions in RMSE are observed in the ranges \(100<A\leq150\) and \(150<A\leq200\), whereas the differences remain small in the other regions.

\begin{table}[!htb]
\renewcommand{\arraystretch}{1.1}
\centering
\caption{RMSE values of the CNN and QCNN in different mass-number ranges.}
\label{tab:rmse_bins_N}
\vspace{1em} 
\setlength{\tabcolsep}{11pt}
\begin{tabular}{lcc}
\hline
\hline
Mass number & CNN (fm) & QCNN (fm) \\
\hline
$0<A\le 50$   & 0.0080 & 0.0088 \\
$50<A\le 100$  & 0.0064 & 0.0064 \\
$100<A\le 150$ & 0.0068 & 0.0046 \\
$150<A\le 200$& 0.0114 & 0.0070 \\
$200<A\le 250$& 0.0077 & 0.0072 \\
\hline
\hline
\end{tabular}
\end{table}

\begin{figure*}
    \centering
    \includegraphics[width=0.95\linewidth]{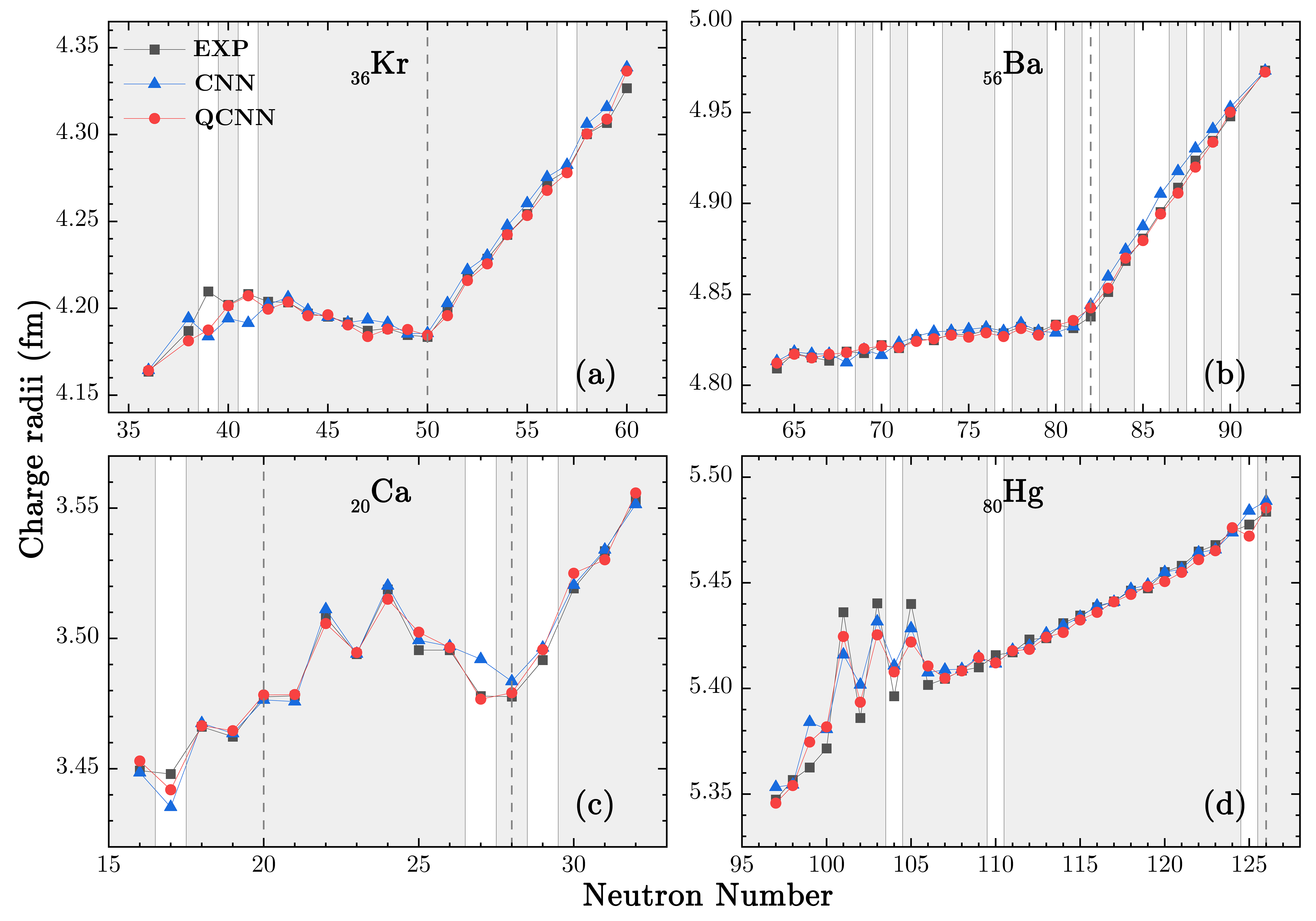}
    \caption{Charge radii predicted by the CNN and QCNN for the Kr, Ba, Ca, and Hg isotopic chains. The vertical dashed lines mark the neutron magic numbers. Experimental data are shown for comparison. The training set is shaded in gray, whereas the validation set is left unshaded.}
    \label{fig:3}
\end{figure*}

We next consider four representative isotopic chains to assess how the models reproduce local nuclear-structure signatures. Figure~\ref{fig:3} compares the experimental charge radii with the CNN and QCNN predictions, while Table~\ref{tab:chain_rmse} lists the corresponding RMSE values. For the Kr and Ba isotopic chains, the charge radii exhibit pronounced kinks around \(N=50\) and \(N=82\), respectively, associated with the corresponding shell closures. Both models reproduce these trends, but the QCNN yields lower RMSEs for both chains. The Ca isotopic chain spans the neutron magic numbers \(N=20\) and \(N=28\) and exhibits a more complex evolution pattern. Along this chain, the QCNN follows the charge-radius evolution more closely than the CNN, particularly for \(^{47}\mathrm{Ca}\). The Hg isotopic chain exhibits pronounced shape staggering in the region \(100\leq N\leq106\). The QCNN also yields a lower chain RMSE for Hg than the CNN.

\begin{table}[!htb]
\renewcommand{\arraystretch}{1.1}
\centering
\caption{RMSE values of the CNN and QCNN along four isotopic chains, together with the neutron-number ranges. For the Ba isotopic chain, the $N=91$ data point is absent.}
\label{tab:chain_rmse}
\vspace{1em} 
\setlength{\tabcolsep}{8pt}
\begin{tabular}{lcc}
\hline
\hline
Isotopic chain & CNN (fm) & QCNN (fm) \\
\hline
Kr ($36\le N\le 60$) & 0.0080 & 0.0055 \\
Ba ($64\le N\le 90, 92$) & 0.0049 & 0.0021 \\
Ca ($16\le N\le 32$) & 0.0052 & 0.0034 \\
Hg ($97\le N\le 126$) & 0.0078 & 0.0068 \\
\hline
\hline
\end{tabular}
\end{table}

\begin{figure*}
    \centering
    \includegraphics[width=0.95\linewidth]{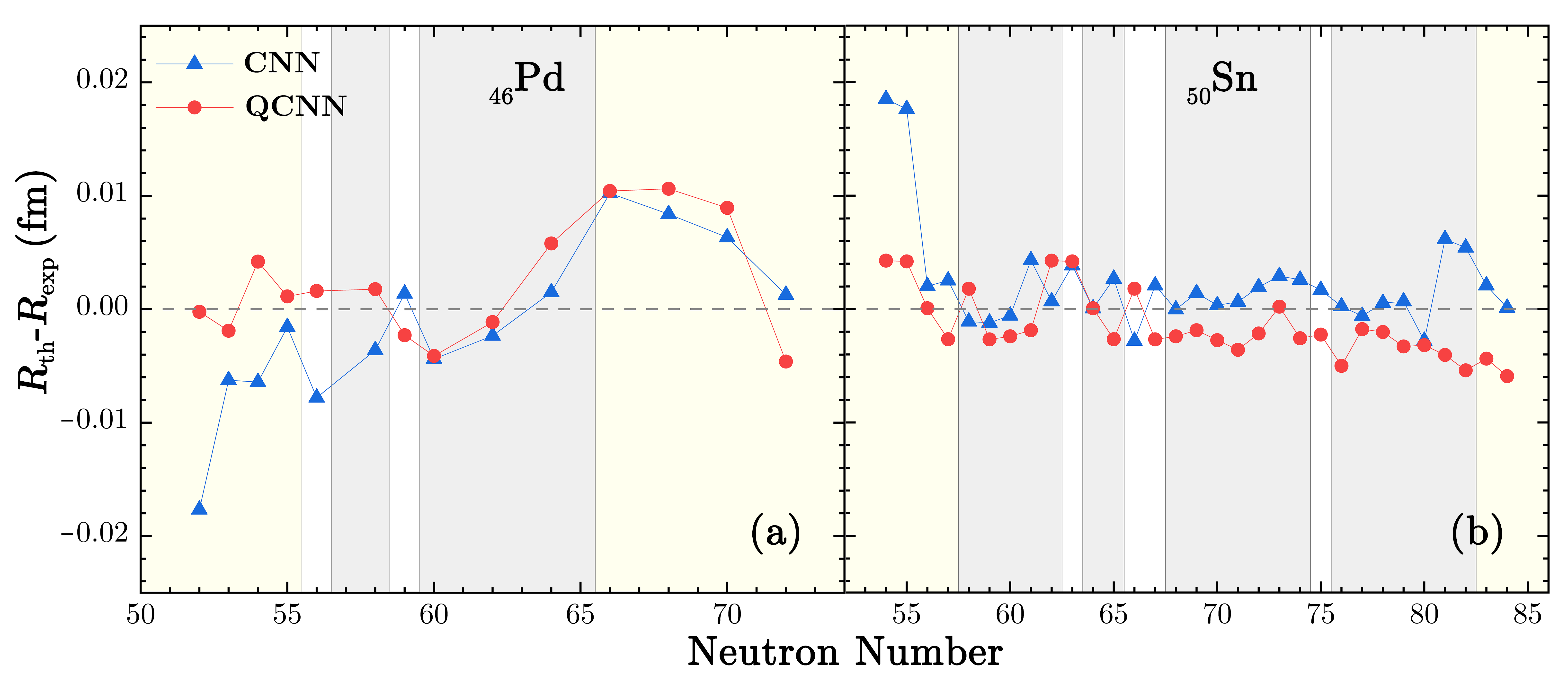}
    \caption{Residuals of the CNN and QCNN predictions relative to the experimental charge radii in the Pd and Sn isotopic chains. Gray shading marks the training set, the white regions mark the validation set, and the yellow shaded regions denote the extrapolation set.}
    \label{fig:4}
\end{figure*}

We then assess performance on the extrapolation set defined in Sec.~\ref{sec:framework}. The CNN and QCNN achieve overall RMSE values of 0.0167 fm and 0.0147 fm, respectively. Thus, the QCNN retains a lower overall error under time-based extrapolation. At the isotopic-chain level, Fig.~\ref{fig:4} presents the residual distributions along the Pd and Sn isotopic chains, which contain the largest numbers of nuclei in the extrapolation set. The residuals of both models remain generally small in the training and validation regions but become larger in the extrapolation regions. More specifically, the QCNN reduces the Pd-chain RMSE from 0.0071 fm to 0.0054 fm and the Sn-chain RMSE from 0.0052 fm to 0.0032 fm.

Finally, to assess robustness against training randomness, each model was trained in 10 independent runs. Across these runs, the mean RMSE values on the full set are 0.0091 fm for the CNN and 0.0071 fm for the QCNN, with standard deviations of 0.0007 fm and 0.0006 fm, respectively. This suggests that the QCNN not only achieves a lower mean error than the CNN but also exhibits smaller fluctuations across repeated runs, indicating better stability. In other words, the lower error of the QCNN is not due to a single favorable training run but is observed consistently across repeated runs.  

The RMSE reduction achieved by the QCNN varies across mass number and is largest in the \(100<A\leq200\) interval. This interval includes nuclei with complex charge-radius systematics associated with shape coexistence, while the selected isotopic chains provide complementary examples involving shell-closure kinks, OES, and shape staggering. Because the two models differ only in the final filter, this pattern suggests that the feature representation is a likely source of the performance difference. The nonlinear multi-input feature map of the variational quantum filter described in Sec.~\ref{sec:framework} offers a plausible mechanism for representing such local variations. The lower errors on the post-2021 set, including the Pd and Sn chains, indicate that the performance difference is partially retained under time-based extrapolation. The smoother validation curve and slightly smaller run-to-run variability also indicate that the lower error is not associated with increased sensitivity to training randomness. Within this controlled comparison, these findings support the interpretation that the variational quantum filter provides a useful representation of complex local charge-radius systematics.

\section{summary and perspectives}
\label{sec:discussion}
In this work, we apply a hybrid QCNN to predict nuclear charge radii and compare it with a classical CNN in a controlled setting. The two models share the same local nuclear-chart input representation, classical convolutional trunk, pooling layer, regression layers, optimizer, and training protocol. The QCNN replaces only the final classical \(2\times2\) convolutional filter with a weight-shared four-qubit variational quantum filter. Under the present data split, the QCNN yields lower RMSE values on the training set, validation set, and full set. Across repeated runs, the QCNN also retains a lower mean full-set RMSE and slightly smaller run-to-run variability. The isotopic-chain analysis shows that the QCNN follows the experimental trends more closely in the selected Kr, Ba, Ca, and Hg chains. For the extrapolation set, the QCNN achieves a lower overall RMSE and lower chain RMSEs for Pd and Sn. Together, these results support the feasibility of the hybrid architecture for nuclear charge-radius prediction under this evaluation protocol.

This study provides a simulation-based test of a NISQ-compatible hybrid architecture for a nuclear-structure task. Rather than relying on a large-scale quantum circuit, the model uses a small reusable quantum filter as a local feature generator embedded in an otherwise classical CNN. Such a design is well suited to problems where local correlations on the nuclear chart are important, and it serves as a practical test platform for exploring whether quantum feature maps can complement classical convolutional structures in nuclear data analysis. While the present calculations use noiseless quantum-circuit simulations,
future work should examine finite-shot sampling and the resulting
stochastic optimization~\cite{kreplin2024reduction,sweke2020stochastic},
realistic hardware noise~\cite{preskill2018quantum,cerezo2021variational},
and error mitigation~\cite{cai2023quantum}. Further studies could also explore richer data encodings~\cite{schuld2021effect}, larger quantum filters, and more physically informed input features. These extensions would help determine whether the performance gains observed in the present simulations remain robust under realistic quantum-computing conditions and whether the approach can scale to more complex nuclear-physics tasks, such as nuclear masses and separation energies~\cite{lunney2003recent}, charge-density distributions~\cite{ravenhall1958electron}, and decay half-lives relevant to nucleosynthesis~\cite{pfutzner2012radioactive}.

Overall, this work provides a preliminary framework for applying quantum convolutional layers to nuclear-structure studies and a controlled setting for further developing hybrid quantum-classical models for nuclear-physics tasks.

\begin{acknowledgments}
We are grateful to Prof. James P. Vary and Dr. Weijie Du for fruitful discussions. This work is supported by the National Natural Science Foundation of China (Grant No. 12475119), the Key Laboratory of Nuclear Data Foundation (JCKY2025201C154), and the JSPS Grant-in-Aid for Scientific Research (S) under Grant No. 20H05648. 
\end{acknowledgments}
\bibliography{apssamp}

\end{document}